\documentclass[sigconf]{acmart}

\AtBeginDocument{%
  \providecommand\BibTeX{{%
    \normalfont B\kern-0.5em{\scshape i\kern-0.25em b}\kern-0.8em\TeX}}}

\setcopyright{ccbysa}
\copyrightyear{2023}
\acmYear{2023}
\acmDOI{}

\acmConference[IMI Workshop'23]{Intelligent Music Interfaces: When Interactive Assistance and Augmentation Meet Musical Instruments}{March 12,
  2023}{Glasgow, UK}
\acmBooktitle{Intelligent Music Interfaces: When Interactive Assistance and Augmentation Meet Musical Instruments, March 12, 2023, Glasgow, UK}
\acmPrice{}
\acmISBN{}

\usepackage{graphicx}
\usepackage{siunitx}

\begin{document}

\title{SolefulTap: Augmenting Tap Dancing Experience \\using a Floor-Type Impact Display
% Inducing a Tap Dancing Experience through Step Augmentation using a Floor-Type Impact Display
}


\author{Tomoya Sasaki}
\authornotemark[1]
\affiliation{
  \institution{The University of Tokyo}
  \state{Tokyo}
  \country{Japan}
  \postcode{153-8904}
}
\email{sasaki@star.rcast.u-tokyo.ac.jp}

\author{Narin Okazaki}
\authornote{Both authors contributed equally to this research.}
\affiliation{%
  \institution{The University of Tokyo}
%   \streetaddress{????}
%   \city{????}
  \state{Tokyo}
  \country{Japan}
  \postcode{153-8904}
}
\email{narin@star.rcast.u-tokyo.ac.jp}

\author{Takatoshi Yoshida}
\affiliation{
  \institution{The University of Tokyo}
  \state{Tokyo}
  \country{Japan}
  \postcode{153-8904}
}
\email{taka_y@star.rcast.u-tokyo.ac.jp}

\author{Alfonso Balandra}
\affiliation{
  \institution{The University of Tokyo}
  \state{Tokyo}
  \country{Japan}
  \postcode{153-8904}
}
\email{alfonso@star.rcast.u-tokyo.ac.jp}

\author{Zendai Kashino}
\affiliation{
  \institution{The University of Tokyo}
  \state{Tokyo}
  \country{Japan}
  \postcode{153-8904}
}
\email{kashino@star.rcast.u-tokyo.ac.jp}

\author{Masahiko Inami}
\affiliation{
  \institution{The University of Tokyo}
  \state{Tokyo}
  \country{Japan}
  \postcode{153-8904}
}
\email{inami@star.rcast.u-tokyo.ac.jp}

\renewcommand{\shortauthors}{Sasaki and Okazaki, et al.}

\renewcommand{\shorttitle}{SolefulTap: Augmenting Tap Dancing Experience using a Floor-Type Impact Display}%

\begin{abstract}
We propose SolefulTap for a novel tap dancing experience.
It allows users to feel as if they are tap dancing or appreciate a tap dancing performance using the sensations of their own feet.
SolefulTap uses a method called Step Augmentation that provides audio-haptic feedback to users, generating impacts in response to users' simple step motions.
Our prototype uses a floor-type impact display consisting of pressure sensors, which detect users' steps, and solenoids, which generate feedback through impact.
Through a preliminary user study, we confirmed that the system can provide untrained users with the experience of tap dancing.
This study serves as a case study that provides insight into how a reactive environment can affect the human capabilities of physical expression and the sensation experienced.
\end{abstract}




\begin{teaserfigure}
  \includegraphics[width=\textwidth]{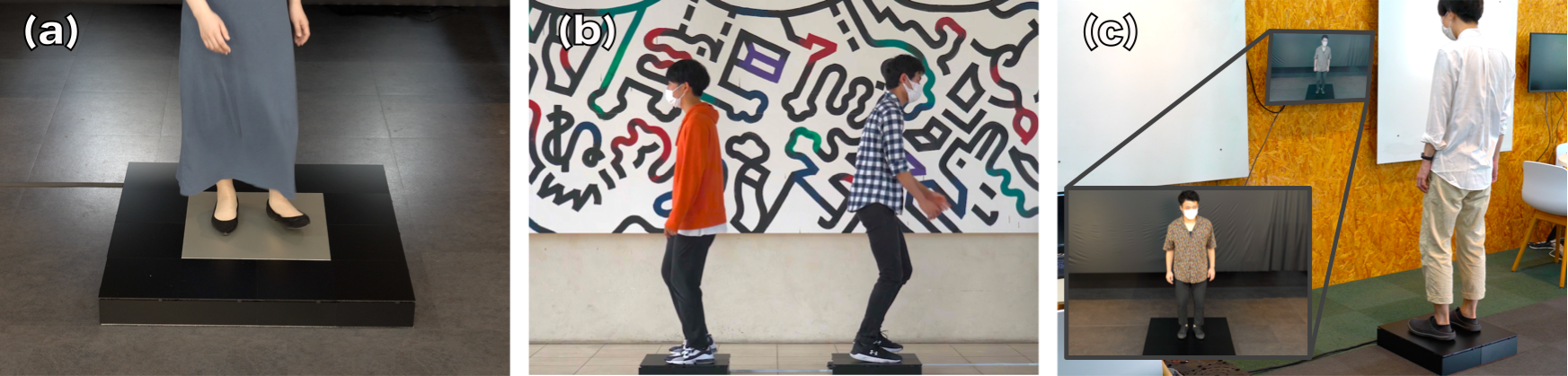}
  \caption{(a) SolefulTap provides tapping sounds and haptic stimulus from a floor-type impact display in response to the user's step motion. (b) Two users can tap dance in unison without directly seeing each other by using the sensation of their soles. (c) By recording the performance data, users can learn tap dancing steps multimodally through visual, audio, and haptic stimuli.}
  \label{fig:teaser}
\end{teaserfigure}


\maketitle

\section{Introduction}
\label{sec:Introduction}
Tap dance is a form of dance expression in which dancers wearing tap shoes tap the floor with their feet and generate sound as a form of percussion.
Through tap dancing, dancers not only express their own body movements visually but also make a beat auditorily by tapping, i.e., interacting with the environment such as the floor.
The steps performed while tap dancing on the floor provide haptic impact sensations to the dancer's sole in addition to sounds.
Thus, tap dancing can be recognized as a dance form in which the dancers express themselves by multimodally utilizing their visual, auditory, and haptic sensations through the interactions between their feet and the floor.
In this study, we aimed at human augmentation during the tap dance experience.

We can distinguish roughly two approaches to augmenting a dance performance: One is to attach devices to the dancers' bodies or props, and the other is to embed a system into the dancers' stage. 
For example, the former uses wearable technology~\cite{Izuta:2019,Karpashevich:2018}, and the latter uses AR technology~\cite{Sparacino:2000,Clay:2012}. These approaches have both advantages and disadvantages depending on the type and composition of the dance.
In the case of tap dancing, a dancer uses special shoes with metal on the soles to tap the floor. 
In other words, it is characterized by the dancer's interaction with the stage using special props worn on the body. 
Augmenting tap dancing would be an interesting subject for research on human augmentation in that it focuses on the interaction of props and environments through body movements.

In this study, we propose ``SolefulTap,'' which provides a novel tap dancing experience by focusing on the interaction between the soles and the floor (Fig.~\ref{fig:teaser}).
This concept uses a method called ``Step Augmentation,'' which generates an impact in response to the dancers' steps on a floor with embedded sensors and actuators.
As a prototype based on the proposed concept, we developed a floor-type device that uses multiple solenoids for generating impacts. 
In addition, we conducted a preliminary user study by demonstrating the prototype.
We explored the possibilities for creating novel dance performances and experiences of viewing them, taking tap dancing as an example by constructing an environment that interacts with users in response to their body movements.
Through this, we discuss the potential of human augmentation based on the interaction between bodies and environments.
\section{SolefulTap}
\label{sec:concept}

SolefulTap provides a novel tap dancing experience by generating impacts that are felt in the soles from the floor in response to users' step motions. The impacts provide both auditory and haptic sensations to users.
We call this method Step Augmentation, which uses a floor-type impact display equipped with sensors and actuators.
Using only the floor-type impact display, the system can detect steps via the user's body movements and present the impact of a tap on the floor to the user's sole.
We can design impact patterns, such as multiple impacts per step input, or impacts to a position different from the user input position.
The impacts can be displayed in real-time, or the input information can be recorded for playback offline.
In addition, users can tap dance without attaching any special devices.
For example, they can tap while wearing sneakers or bare feet.
By designing the impact pattern of Step Augmentation for each application, SolefulTap provides a variety of novel tap dancing experiences.

\begin{figure}[tbp]
    \centering
    \includegraphics[width=0.48\textwidth]{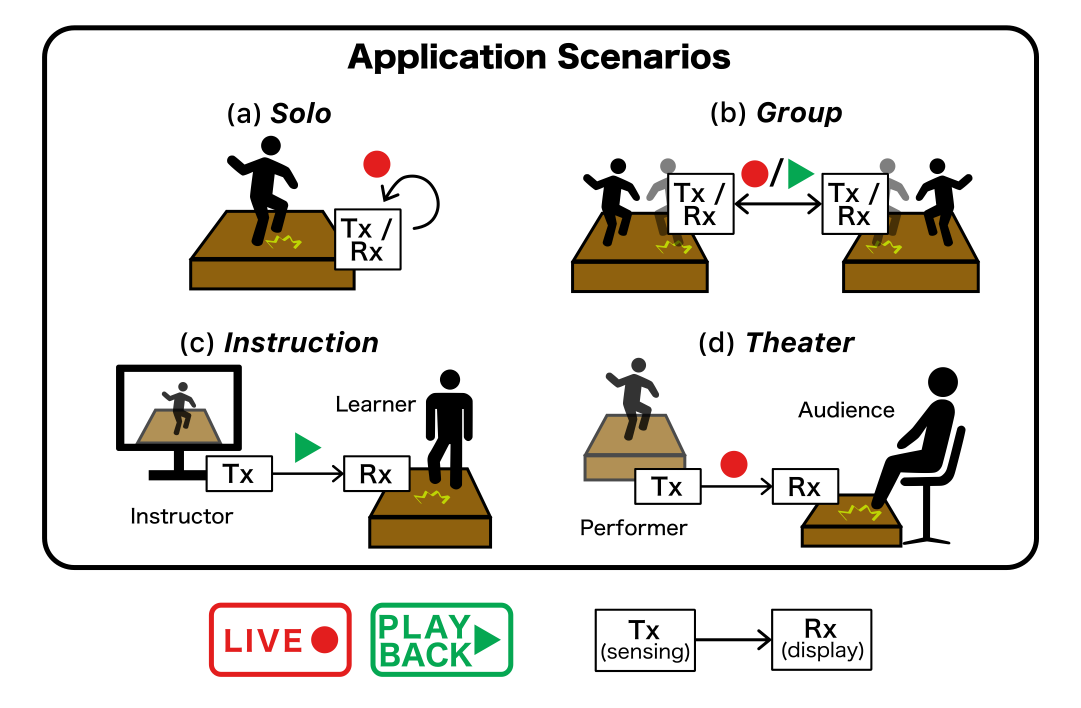}
    \caption{Four application scenarios of SolefulTap: (a)~\textit{\textsf{Solo}}, (b)~\textit{\textsf{Group}}, (c)~\textit{\textsf{Instruction}}, and (d)~\textit{\textsf{Theater}}. 
    Information can be transmitted from Tx to Rx in real-time, or the recorded data can be played back offline.}
    \label{fig:DesignSpace}
\end{figure}

SolefulTap provides several possible application scenarios for a tap dancing experience according to the number of users and the combination of impact patterns.
We considered the cases of one to multiple users by categorizing a user as a performer or audience.
Herein, we illustrate four specific application scenarios, \textit{\textsf{Solo}, \textsf{Group}, \textsf{Instruction}}, and \textit{\textsf{Theater}} (Fig.~\ref{fig:DesignSpace}).

\textit{\textsf{Solo}} --- 
This is the basic form of SolefulTap, which generates impacts on the soles in response to the steps of a single performer in real-time (Fig.~\ref{fig:DesignSpace}~(a)).
By setting up multiple impact patterns in advance, it is possible to generate a variety of tapping sounds and haptic stimulations with a single step.
This I/O coincidence interaction design allows performers to express a tap dance routine as if the impacts are part of the performer’s own steps despite actually coming from the device (Fig.~\ref{fig:teaser}~(a)).

\textit{\textsf{Group}} --- 
Using multiple devices, a performer can tap dance with a distant partner by transmitting information bidirectionally in real-time (Fig.~\ref{fig:DesignSpace}~(b)). 
For example, when two performers tap dance together, they can feel their partner's steps through the auditory and haptic sensations of the impact display, which makes it possible to dance in unison without the dancers directly seeing each other (Fig.~\ref{fig:teaser}~(b)).

\textit{\textsf{Instruction}} --- 
By recording the steps of an instructor who is an experienced performer, and playing the recording back offline along with other video images, beginners can learn to tap dance (Fig.~\ref{fig:DesignSpace}~(c)).
There are many step variations in tap dance, and it is difficult to understand the correct step timing and position by simply playing a video because the steps are performed rapidly.
By generating the impacts from the points corresponding to the instructor's step positions, learners can perceive the steps multimodally.
In addition, it is possible to take advantage of the valuable functions of the recorded data such as slow playback and rewinding (Fig.~\ref{fig:teaser}~(c)).

\textit{\textsf{Theater}} --- 
By transmitting the performers' steps, which are detected on the stage, to multiple devices for viewing by the audience, the system can be applied as a tap dance appreciation tool (Fig.~\ref{fig:DesignSpace}~(d)).
In general, audiences can only enjoy a performance audiovisually if the stage is at a distance.
With SolefulTap, they can also feel a haptic sensation at their own feet in response to the performance.
In addition, by using multiple devices to show the impacts of the steps of a single performer, amplification can be achieved, simulating numerous performers tap dancing at the same time, despite the dance being performed by a single dancer.

% In the next section, we describe the implementation of a prototype system for a floor-type device that displays the impacts based on the \textit{\textsf{Solo}} scenario.
\begin{figure*}[tbp]
\captionsetup{justification=centering}
 \begin{center}
  \begin{minipage}[c]{0.29\linewidth}
    \centering
    \includegraphics[width=0.9\hsize]{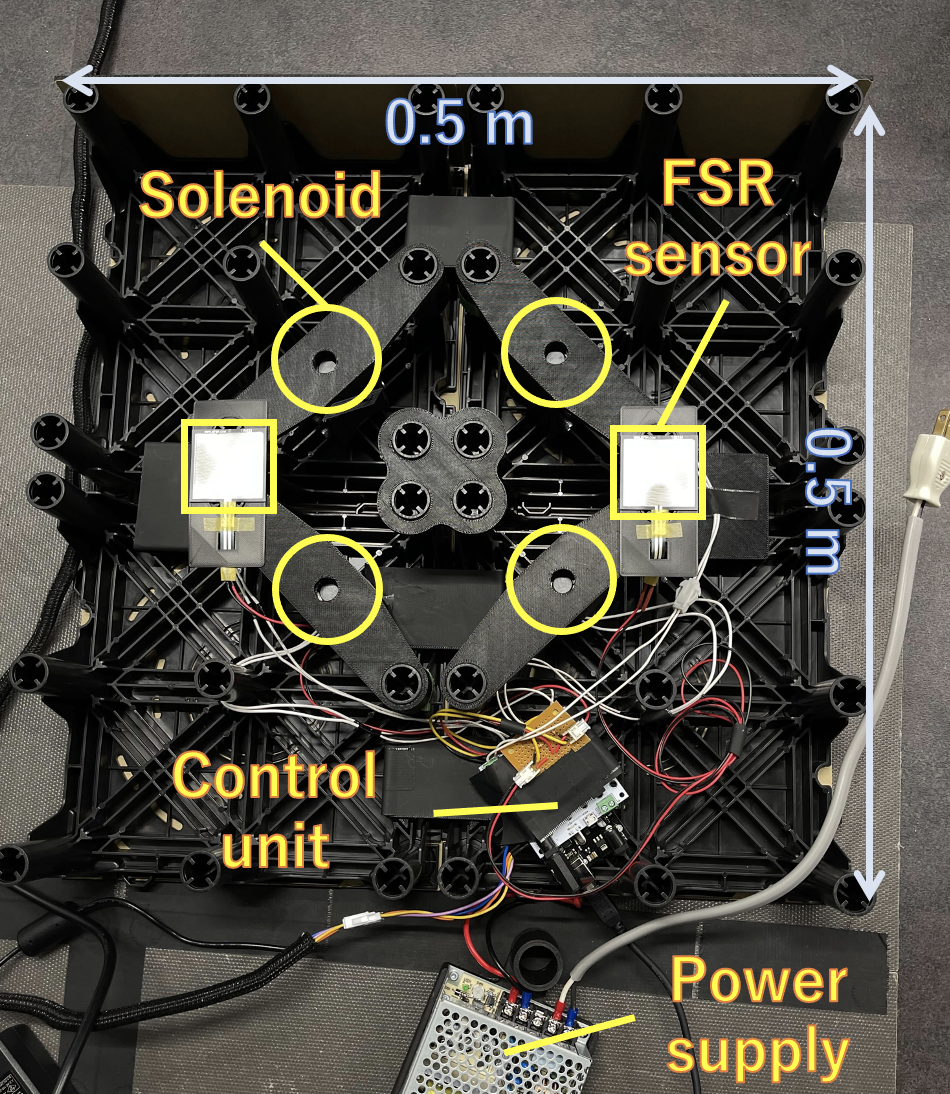}
    \caption{Hardware components and dimensions of the device (back side of the device)}
    \label{fig:hardware}
  \end{minipage}
  \begin{minipage}[c]{0.7\linewidth}
    \centering
    \includegraphics[width=1.0\hsize]{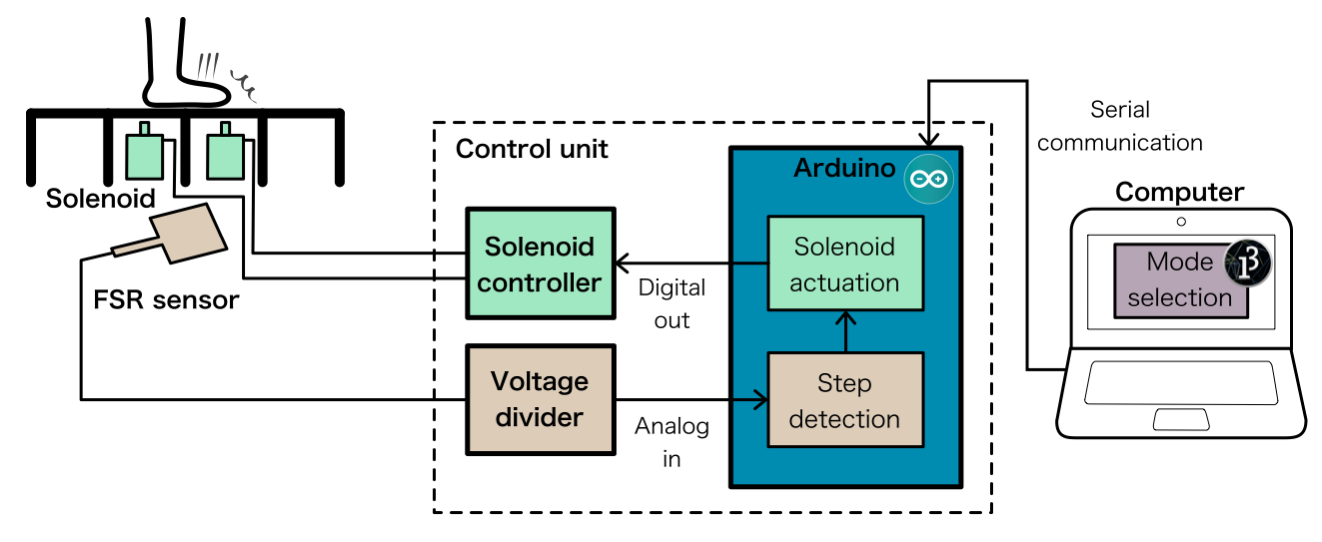}
    \caption{Schematic diagram of the system configuration \\(The device part is shown as a right-side view.)}
    \label{fig:SystemConfiguration}
  \end{minipage}
 \end{center}
\end{figure*}

\section{Implementation}
\label{sec:SystemDesign}
We introduce the prototype floor-type device for Step Augmentation, in order of hardware module and control software.

% \subsection{Hardware Module}
\label{sec:HardwareModule}
We designed the floor-type apparatus as a module.
The device frame is laid floor tiles (T-100R, Fukuvi Chemical Industry Co., Ltd.).
The dimensions of each tile are \SI{0.5}{m}~(W) $\times$ \SI{0.5}{m}~(D) $\times$ \SI{0.1}{m}~(H).
Pressure sensors and solenoids were used for step detection and impact display, respectively, and were installed at the bottom of the device, as shown in Fig.~\ref{fig:hardware}.
The four solenoids (CB12500770, Takaha Kiko Co., Ltd.) were controlled by a driver circuit (SB-6565-01, Takaha Kiko Co., Ltd.) with \SI{24}{V} applied by a switching power supply.
Two pressure sensors (FSR406, Interlink Electronics Inc.) were installed, one each on the left and right sides.
The resistance of the voltage divider circuit connected to the pressure sensor was set to \SI{1}{k\ohm} to prevent the saturation of the sensor value.
A microcontroller (Arduino Uno) was used to control the sensors and driver and to interface with the PC.

% \subsection{Control Software}
\label{sec:ControlSoftware}
% system config
A schematic diagram of the system configuration from step detection to impact display is shown in Fig.~\ref{fig:SystemConfiguration}.
The pressure sensors detect the user's steps, whereas the solenoids are controlled to generate impact patterns designed for each application.
For this prototype, we implemented a \textit{\textsf{Solo}} mode and a \textit{\textsf{Theater}} mode based on the application scenario described in Section~\ref{sec:concept}.
The PC and Arduino are connected through serial communication, and the modes can be switched from the PC.

% sensing algorithm
The pressure sensor values used for step detection are read as analog inputs at a sample rate of \SI{5}{ms}.
The difference between subsequent sensor readings was calculated as $\delta_{t}(=S_{t}-S_{t-1})$, where $S_t$ and $S_{t-1}$ are the actual and previous sensor values, respectively.
We then used $\delta_{t}$ as the input of a z-score-based method~\cite{stackoverflow-PeakDetection} to detect the signal peaks of the pressure sensors.
The same algorithm was applied to detect both left and right sensor input and determine which foot the user used to perform the step.

% solenoid actuation rule
The impact display is achieved by actuating the solenoids on the corresponding side of the stepped floor.
The driver is controlled by the digital output (ON/OFF) from the Arduino, and the solenoid corresponding to the digital pin whose output is ON is actuated.
The number of impacts by the solenoids and the impact display module can be specified by selecting the solenoid actuation rules on the PC.
For our prototype, the number of impacts by the solenoids was set to one, two, or three.
In the case of a single impact, the solenoid located at the front side is actuated once when the user takes a step.
In the case of two impacts, the solenoids are actuated one at a time in order of front to back, and in the case of three impacts, the solenoids are actuated one at a time in order of front to back to front. %ここ分かりにくいかも
The time interval between multiple impacts per step was set to \SI{90}{ms}.

\section{Preliminary User Study}
\label{sec:PreliminaryUserStudy}
We evaluated the performance of SolefulTap through several demonstrations as a preliminary user study.
Each demonstration was shown to 13 participants (including one person with tap dancing experience), and we obtained their verbal feedback. 
Among the participants, four males and two females (aged 22--41, with an average age of 27) also responded to a questionnaire survey.
All participants first experienced the \textit{\textsf{Solo}} scenario, and then were then asked to freely try the other scenarios, including impact pattern selection.

First, we report the results of the questionnaire. 
We asked the participants to choose two options (``felt'' or ``did not feel'') to the question, ``Did you feel any sense of incongruity in the position of the impacts you were given?'' All of the participants answered ``no.'' 
This shows that the coincidence of the impact positions for the left and right steps and the response time are sufficient.
To the question ``Did you feel that you were tap dancing?'', the participants were asked to answer on a 7-point Likert scale (1 = not at all, 7 = very much), with 83.5\% of them answering 4 or more (mean of 5, median of 5.5). 
Next, we describe the impressions and comments obtained verbally.
Regarding the mechanism of the impact display itself, one participant commented that ``it is like tap dancing,'' after trying the device for the first time without being instructed on the concept of SolefulTap beforehand.
Regarding the number of impacts per step, many participants commented that the experience was more enjoyable and that they felt more like tap dancing when two or three impacts were given rather than just one.
From the above results of the questionnaire, it was suggested that the auditory and haptic sensations produced by the impact display of the prototype made the participants recall their own experience with tap dance.

In addition to the above, there were two interesting topics among the comments obtained through this demonstration.
The first is the issue of variations in the expression in the prototype, as commented by the experienced participant. 
The strength and tone of the tap sound are a part of the expression in tap dancing, and it is necessary to reflect the intention of the performer.
Therefore, in addition to the number of impacts by the solenoids, parameters of the solenoid actuation pattern, such as the adjustment of the intensity, should be added.
Moreover, the sensors should accurately detect the steps.
A control system is required to respond to user inputs, such as a light or gradual change in the step force.
It is possible for performers to finely control the impact pattern depending on the intensity and rhythm of the step, which will contribute to a broader variety of expressions.
The second issue was based on a comment from a participant regarding the sense of body shift.
The participant said that when switching to the \textit{\textsf{Theater}} mode after experiencing the \textit{\textsf{Solo}} mode, it felt as if a part of the participant's body had shifted.
When switching modes, we stopped providing the impacts that the participant had been given in the \textit{\textsf{Solo}} mode and provided the impacts to a module on the audience side that was within visual and audible distance.
It is thought that this caused a sense of loss of feedback regarding the participant's own stepping movements, and at the same time, the participant perceived the impact in a different position, which caused a feeling that the participant’s body had shifted.

In summary, from the preliminary user study, we confirmed that the proposed method works properly and recalls the experience with tap dance for those who have not experienced it.
However, the experienced participant commented on the variety of expressions.
In addition, it was suggested that Step Augmentation can provide a sense of body shift that does not occur in normal tap-dancing performances.

\section{Discussion}
% \section{Discussion and Conclusion}
\label{sec:DiscussionAndFutureWork}
Based on the prototyping of SolefulTap and the findings of the preliminary user study, we will now discuss the design of Step Augmentation, its scalability, and our contributions to human augmentation.

% design of Step Augmentation
First, regarding the design of Step Augmentation, we confirmed that displaying multiple successive impacts for a single step input provided the experience of tap dancing for inexperienced users.
However, this design has limitations in terms of providing flexible expressions for experienced users.
Thus, it was suggested that the benefits of Step Augmentation in the \textit{\textsf{Solo}} mode be different for inexperienced and experienced users.
Regarding the exploration of novel expressions in tap dance, the following two future research directions can be considered for experienced dancers.
First, a person who has no experience in tap dancing but has experience in different dance forms such as street dance can use the \textit{\textsf{Solo}} mode to combine the expression of two dance techniques.
Second, experienced tap dancers can use the \textit{\textsf{Group}} or \textit{\textsf{Theater}} mode to amplify their own expression.
Both of these directions can be approached by improving the design of Step Augmentation and the floor-type display, without restricting the dancers' own physical expressions.

\begin{comment}
% sense of agency
Next, we discuss the sense of tap dancing from the perspective of the sense of agency under temporal and spatial factors.
It is important to have a sense of agency while tap dancing, allowing one to feel that the tap is being generated by one's own motion, even when the technology is intervening.
For the temporal factor, it is generally known that the sense of agency is lost when the time delay of the feedback to one's own action is approximate \SI{200}{ms} or longer~\cite{Engbert:2008}.
We confirmed that the time delay between the user input and the first impact by a solenoid was sufficiently short, i.e., \SI{30}{ms}, in our prototype, which satisfies this requirement. 
The result of the users' feedback in the questionnaire also supports this fact.
As for the spatial factors, it was suggested that the sense of agency may arise even when the positions of the input and output do not coincide, according to a report indicating that a sense of body shift was provoked when switching from the \textit{\textsf{Solo}} to \textit{\textsf{Theater}} mode.
The change in the sense of self-location when switching modes may affect the dance expression.
How these changes affect dancers' expressions needs to be investigated in the future.
\end{comment}

% scalability
Next, regarding the scalability of SolefulTap, the proposed system is scalable because it is modularized, as shown in the prototype.
It can be applied to large areas, such as a stage or an auditorium in a concert hall by laying out numerous modules, as Yoshida et al.~\cite{Yoshida2022-nv} proposed a modular floor sensing system using floor tiles.
Using it on a stage, more diverse dance expressions can be achieved, such as by moving around in a large space or by performing a dance simultaneously with several people.
In an auditorium, many audience members can enjoy the performance in \textit{\textsf{Theater}} mode.
It can also be combined with Shibasaki et al.'s approach~\cite{Shibasaki:2016} or other wearable technologies to augment the dance expressions.
We believe that these aspects will allow us to explore even more diverse novel tap dance experiences.

% human augmentation
Finally, as the contributions of this research to human augmentation, there is a notion that human capabilities do not belong to individuals but are formed in the context of relationships with other people and environments.
For example, regarding motility, the philosopher Merleau-Ponty stated that ``In order that we may be able to move our body towards an object, the object must ﬁrst exist for it''~\cite{merleau1962phenomenology}.
That is, not only individual will, but also environmental factors exist for human motility.
The system proposed in this study can be abstracted as ``an environment that expands the possibility of human self-expression and reconstructs relationships with others through the augmentation of body motion.''
Dance is a physical activity showcasing human expressions, and our system attempts to augment such expressions without attaching any device to the dancer.
We believe that providing a concrete example of an approach to augmentation from an environment will provide insight into future human augmentation research.

\section{Conclusion}
% In conclusion, 
We proposed SolefulTap, which provides a novel tap dancing experience through Step Augmentation using a floor-type impact display.
Step Augmentation is achieved by providing audio-haptic feedback to users in response to their steps.
The feedback is generated by impacting the floor surface with a set of solenoids.
% We explored user interactions and experiences created by the system through prototype demonstrations, representing application scenarios.
% After each demonstration, a user survey was conducted using a questionnaire, and verbal feedback was received.
% As a result, we confirmed that the system was able to provide untrained users with the experience of tap dancing.
% Given the findings, we discussed the design of impact patterns and the scalability of the system as well as the future prospects of applying the proposed system.
SolefulTap also represents an approach to human augmentation that focuses on the interaction between human bodies and environments, specifically the soles and floor.
Despite not wearing any device on the user’s body, our approach provides a physical expression that feels as if the user is performing a tap dance based on reactions from the environment.
Our study, therefore, is an exploration of novel forms of physical expression through interactions with a reactive environment.
We hope that our viewpoint will help diversify the approaches toward building human capabilities in the field of human augmentation.
% \input{sections/conclusion.tex}

\begin{acks}
This project is supported by JST ERATO JIZAI BODY Project (JPMJER1701), Japan.
\end{acks}

\bibliographystyle{ACM-Reference-Format}
\bibliography{reference}

\end{document}